 \documentclass[preprint,prd,tightenlines, superscriptaddress]{revtex4-1}

\usepackage{graphicx} 
\usepackage{dcolumn}  
\usepackage{colordvi}
\usepackage{color}
\usepackage{epstopdf}
\usepackage{pstricks}
\usepackage{amssymb}
\usepackage{url}
\graphicspath{{ps}}
\usepackage{hyperref}
\usepackage{tabularx}
\usepackage{multirow}
\usepackage{units}
\usepackage{upgreek}
\usepackage{siunitx}
\usepackage{hyphenat}
\usepackage{subfigure}
\usepackage[italic]{hepnames}

\newcommand{\CP}{\ensuremath{C\hspace{-0.13em}P}\xspace}

\renewcommand{\Ph}{\HepParticle{h}{}{}\xspace}

\renewcommand{\Prhozero}{\ensuremath{\HepParticle{\Prho}{}{}^0}\xspace}

\renewcommand{\APDzero}{\ensuremath{\HepParticle{\APD}{}{}^0}\xspace}
\renewcommand{\Pgpz}{\ensuremath{\HepParticle{\Pgp}{}{}^0}\xspace}
\renewcommand{\PDzero}{\ensuremath{\HepParticle{\PD}{}{}^0}\xspace}

\begin{document}

\def\belletwo {\it {Belle~II}}

\clubpenalty = 10000  
\widowpenalty = 10000 

\vspace*{-3\baselineskip}
\resizebox{!}{3cm}{\includegraphics{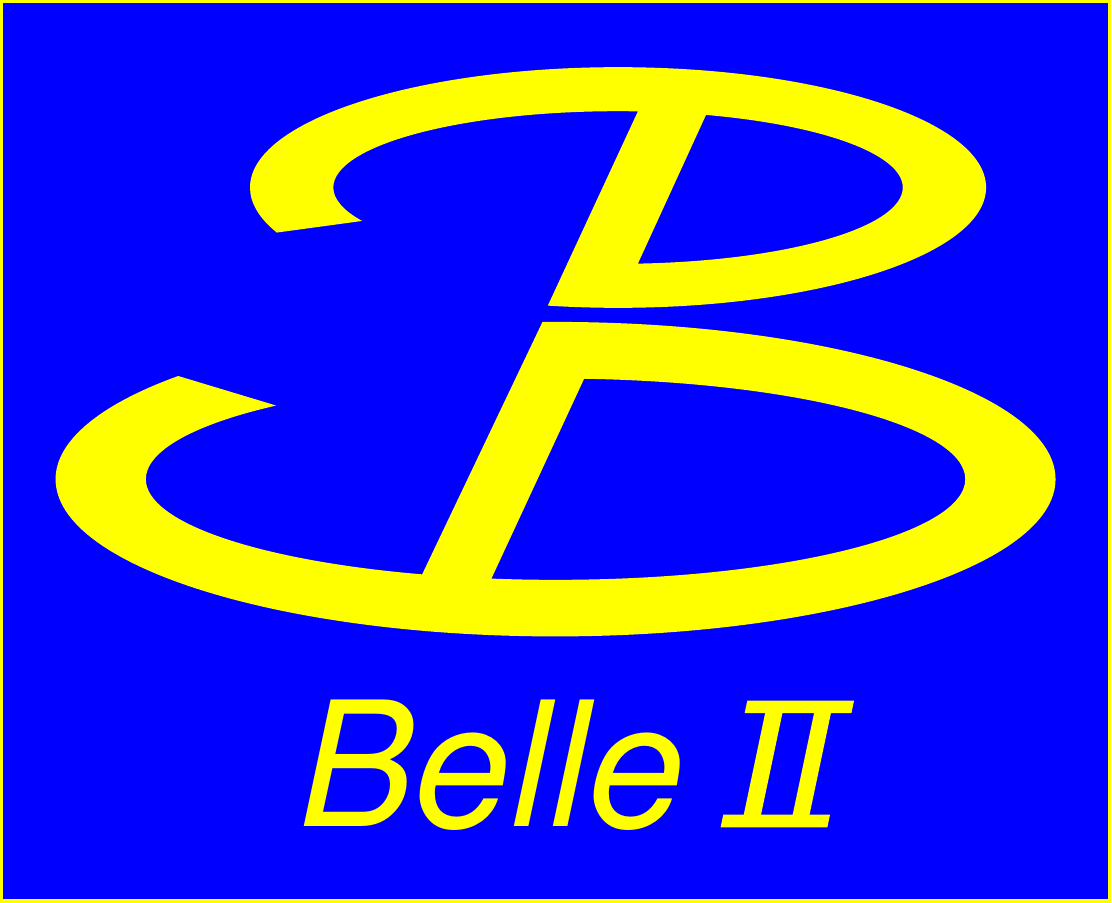}}

\vspace*{-5\baselineskip}
\begin{flushright}
BELLE2-CONF-PH-2022-005

\today
\end{flushright}

\quad\\[0.5cm]

\title {Angular analysis of $B^+ \to \rho^+\rho^0$ decays reconstructed in 2019, 2020, and 2021 Belle~II data}

\collaboration{The Belle II Collaboration}
  \author{F. Abudin{\'e}n}
  \author{I. Adachi}
  \author{K. Adamczyk}
  \author{L. Aggarwal}
  \author{P. Ahlburg}
  \author{H. Ahmed}
  \author{J. K. Ahn}
  \author{H. Aihara}
  \author{N. Akopov}
  \author{A. Aloisio}
% ...
  
\begin{abstract}
We report on a Belle II measurement of the branching fraction~($\mathcal{B}$), longitudinal polarization fraction ~($f_L$), and \CP~asymmetry~($\mathcal{A}_{\CP}$) of $B^+\to \rho^+\rho^0$ decays.  We reconstruct $B^+\to \rho^+(\to \pi^+\pi^0(\to \gamma\gamma))\rho^0(\to \pi^+\pi^-)$ decays in a sample of SuperKEKB electron-positron collisions collected by the Belle II experiment in 2019, 2020, and 2021 at the $\Upsilon$(4S) resonance and corresponding to 190\,fb$^{-1}$ of integrated luminosity.  We fit the distributions of the difference between expected and observed $B$ candidate energy, continuum-suppression discriminant, dipion masses, and decay angles of the selected samples, to determine a signal yield of $345 \pm 31$ events. The signal yields are corrected for efficiencies determined from simulation and control data samples to obtain 
\begin{center}
$\mathcal{B}(B^+ \to \rho^+\rho^0) = [23.2^{+\ 2.2}_{-\ 2.1} (\rm stat) \pm   2.7 (\rm syst)]\times 10^{-6}$,
\end{center}
\begin{center}
$f_L = 0.943 ^{+\ 0.035}_{-\ 0.033} (\rm stat)\pm 0.027(\rm syst)$,
\end{center}
\begin{center}
$\mathcal{A}_{\CP}=-0.069 \pm 0.068(\rm stat) \pm 0.060 (\rm syst)$.
\end{center}
The results agree  with previous measurements. This is the first measurement of $\mathcal{A}_{\CP}$ in $\PBplus\to \Prhoplus\Prhozero$ decays reported by Belle II.
 
\end{abstract}


\maketitle

{\renewcommand{\thefootnote}{\fnsymbol{footnote}}}
\setcounter{footnote}{0}



\section{Introduction and motivation}

\begin{figure}[h!]
 \centering
    \centering
    \includegraphics[width=0.5\textwidth]{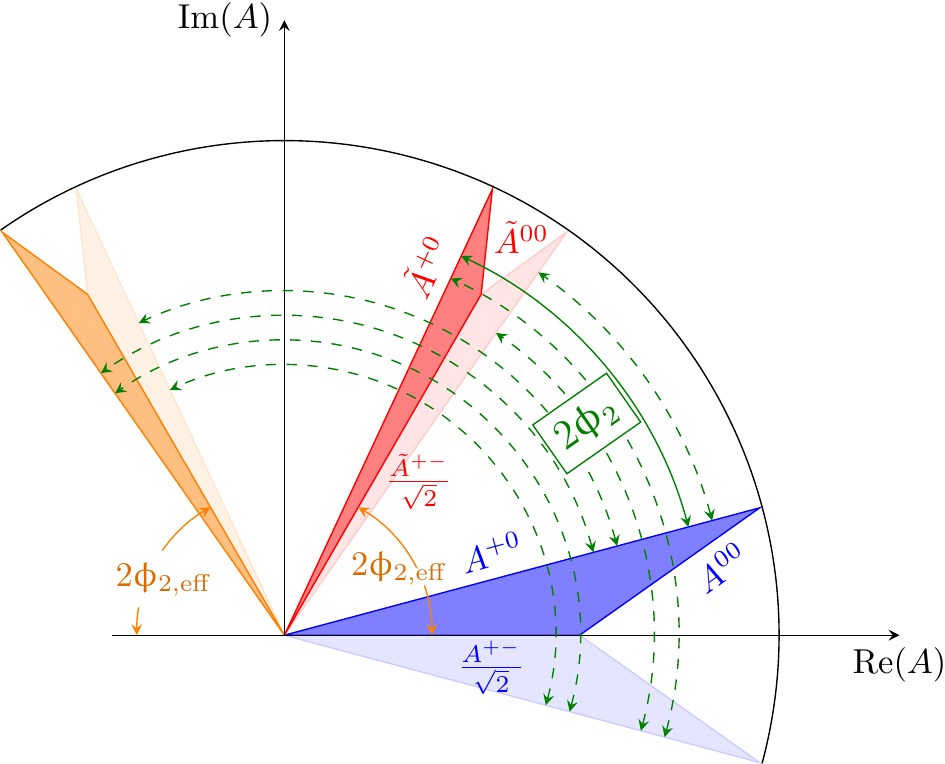}
 \caption{Geometrical representation of the isospin triangular relations~\cite{Gronau:1990ka} in the complex plane of $\PB^{i+j}\to \Ph^i\Ph^j$ amplitudes. The blue and red shaded regions correspond to the isospin triangles. The angle between the \CP-conjugate amplitudes $A^{+-}$ and $\tilde{A}^{+-}$ corresponds to twice the weak phase $\phi_{2,{\rm eff}}$ (orange solid lines). The angle between the \CP conjugate amplitudes $A^{+0}$ and $\tilde{A}^{+0}$ corresponds to twice the quark-mixing angle $\phi_2$ (green solid line). The other triangles with lighter shade represent the mirror solutions allowed by discrete ambiguities
 , with the corresponding values for $\phi_2$ represented by the green dashed lines.}
 \label{fig:isospin_relations}
\end{figure}

The study of charmless $B$ decays is a keystone of the worldwide flavor program. Processes mediated by $b\to u\bar{u}d$ transitions offer direct access to the parameter \mbox{$\phi_2/\alpha \equiv \arg \left[\frac{-V_{td}V^*_{tb}}{V_{ud}V^*_{ub}}\right]$}, where $V_{ij}$ are elements of the quark-mixing matrix. In addition, they probe contributions of non-standard-model dynamics to processes involving internal exchanges of $W$ bosons or heavy quarks (loops)~\cite{Kou:2018nap}.  However, a  reliable extraction of weak phases and an unambiguous interpretation of measurements involving loop amplitudes in non-leptonic $B$ decays are spoiled by hadronic uncertainties,  which often cannot be precisely determined in perturbative  calculations. Appropriately chosen combinations of measurements from decay channels related by flavor symmetries are used to reduce the impact of such  unknowns. An especially  successful approach consists in combining measurements of decays related by isospin symmetries. In particular, the combined analysis of branching fractions and charge-parity (\CP) violating asymmetries of the complete set of   $B \to \rho\rho$ isospin partners  enables a determination of~$\phi_2$~\cite{Gronau:1990ka} (Fig.~\ref{fig:isospin_relations}). Belle~II, which has the  {\it unique}  capability  of  studying  jointly,  and within the same experimental environment, all relevant final states, offers promising opportunities for determining $\phi_2$.

This document describes an improved measurement of branching fraction and fraction of longitudinally polarized of $\PBplus\to \Prhoplus\Prhozero$ decays. Both are essential inputs for the determination of $\phi_2$, along with a 
first measurement of the direct \CP~asymmetry~$\mathcal{A}_{\CP}$. Signal decays are reconstructed in their $\Pgpp\Pgpz(\to\gamma\gamma)\Pgpp\Pgpm$ final state. Charge-conjugate processes are implied.\par

The Belle~II detector, complete with its vertex detector, started its colliding beam \mbox{operations} in March 2019. 
We use the sample of electron-positron collisions collected through June 2021, corresponding to an integrated luminosity of 190\,fb$^{-1}$ collected at the $\Upsilon(4{\rm S})$ resonance and 18\,fb$^{-1}$ collected at an energy about 60\,MeV lower.
All analysis choices and procedures are developed and finalized in simulated and control-sample data before examining signal-rich data.\par Pions in the final state and the large width of $\rho$ mesons reduce distinctive features against dominant backgrounds from $e^+e^- \to q\bar{q}$ events (continuum), where $q$ indicates any quark of the first or second family. Isolating a significant, low-background signal is therefore the main challenge of the analysis. We achieve this through an optimization of the selection requirements followed by a sample-composition fit, both based on the following observables:
\begin{itemize}
    \item the beam-energy-constrained mass $M_{\rm bc} \equiv \sqrt{(\sqrt{s}/2)^{2}- (p^{*}_B)^2}$, which is the invariant mass of the $B$ candidate where its energy is replaced by the (more precisely known) half of the center-of-mass collision energy and $p^*_B$ is the $B$ candidate momentum in the $\Upsilon$(4S) frame, discriminates fully reconstructed $B$ decays from continuum;
    \item the energy difference $\Delta E \equiv E^{*}_{B} - \sqrt{s}/2$ between the energy of the reconstructed $B$ candidate and half of the collision energy, both in the $\Upsilon$(4S) frame, discriminates between signal and misreconstructed $B$ decays and continuum;
    \item dipion masses for $\Prhoplus\,(\to\pi^+\pi^0)$ and $\Prhozero\,(\to\pi^+\pi^-)$ candidates, offer further discrimination against nonresonant and misreconstructed $B$ decays;
    \item the cosines of the helicity angles between the momentum of the positive-charge pion and the direction opposite to the $B^+$ momentum as measured in the $\rho$ rest frame, provide information on the orbital angular momentum of the final state, which can take values of 0, 1, or 2  due to a spin-0 particle decaying into two spin-1 particles. 
\end{itemize}

\section{The Belle~II detector}
Belle~II is a nearly $4\pi$ particle-physics spectrometer~\cite{Kou:2018nap, Abe:2010sj}, designed to reconstruct the products of electron-positron collisions produced by the SuperKEKB asymmetric-energy collider~\cite{Akai:2018mbz}, located at the KEK laboratory in Tsukuba, Japan. Belle~II comprises several subdetectors arranged around the interaction space-point in a cylindrical geometry. The innermost one is the vertex detector, which uses position-sensitive silicon layers to sample the trajectories of charged particles (tracks) in the vicinity of the interaction region to extrapolate the decay positions of their long-lived parent particles. The vertex detector includes two inner layers of silicon pixel sensors and four outer layers of silicon microstrip sensors. The second pixel layer is currently incomplete and covers only one sixth of the azimuthal angle. Charged-particle momenta and charges are measured by a large-radius, helium-ethane, small-cell central drift chamber, which also offers charged-particle-identification information (PID) through a measurement of particles' energy-loss by specific ionization. A  time-of-propagation Cherenkov detector surrounding the chamber provides PID in the central detector volume, supplemented by proximity-focusing, aerogel, ring-imaging Cherenkov detectors in the forward region. A CsI(Tl)-crystal electromagnetic calorimeter allows for energy measurements of electrons and photons.  A solenoid surrounding the calorimeter generates a uniform axial 1.5\,T magnetic field. Layers of plastic scintillators and resistive-plate chambers, interspersed between the
magnetic flux-return iron plates, allow for identification of $K^0_{\rm L}$ and muons.
The subdetectors most relevant for this work are the vertex detector, drift chamber, the PID detectors, and the electromagnetic calorimeter.

\section{Selection and reconstruction}
\label{sec:selection}

We reconstruct the two-body decay $\PBplus\to \Prhoplus\,(\to\pi^+\pi^0)\,\Prhozero\,(\to\pi^+\pi^-)\,$ and the control channels
$B^+ \to \overline{D}^0 (\to K^+ \pi^-)\, \rho^+\,(\to\pi^+\pi^0)$, for analysis validation; $B^+ \to \overline{D}^0 (\to K^+ \pi^- \pi^0)\, \pi^+$,  for validation of continuum-suppression variables and optimization of the $\pi^0$ selection; \mbox{$B^0 \to D^{*-} (\to \overline{D}^0 (\to K^+ \pi^- \pi^0)\, \pi^-)\pi^+$} and \mbox{$B^0 \to D^{*-} ( \to \overline{D}^0 (\to K^+ \pi^-)\, \pi^-)\pi^+$}, to determine the systematic uncertainty associated with the $\pi^0$ reconstruction efficiency;
$B^+ \to \overline{D}^0 (\to K^+ \pi^- )\, \pi^+$, to determine the systematic uncertainty associated with the continuum-suppression and PID; and \mbox{$D^{+} \to K_S^0 \pi^+$}, to pinpoint any charge asymmetry due to charge-dependent differences in interaction probabilities, track reconstruction (tracking), or PID.

\subsection{Simulated and experimental data}
We use signal-only simulated data to model relevant signal features for the sample-composition fit and determine selection efficiencies. We use generic simulated data to optimize the event selection and construct the sample-composition fit model for backgrounds.
The generic simulation consists of Monte Carlo samples that include $e^+e^-\to B^0\overline{B}^0$, $B^+B^-$, $u\bar{u}$, $d\bar{d}$, $c\bar{c}$, and $s\bar{s}$ processes in realistic proportions and corresponding in size to six times the $\Upsilon$(4S)~data. In addition, $10\times 10^6$ signal-only $B^+\to\rho^+\rho^0$ events are generated for both polarizations (longitudinal and transverse)~\cite{Lange:2001}, and $6\times 10^4$ events are generated for leading peaking background sources. Finally, simplified simulated experiments constructed by randomly sampling the likelihood of the sample-composition fit allow for studying the estimator properties and assessing fit-related systematic uncertainties. \par 
We use all 2019--2021  $\Upsilon$(4S) good-quality experimental data collected through June 2021, which correspond to an integrated luminosity of $190\,\si{fb^{-1}}$. All events are required to satisfy loose selection criteria, based on total energy and charged-particle multiplicity in the event, targeted at reducing sample sizes to a manageable level with negligible impact on signal efficiency. All data are processed with the Belle~II analysis software~\cite{Kuhr:2018lps}.

\subsection{Reconstruction and baseline selection}
We form final-state particle candidates by applying loose baseline selection criteria and then combine candidates in kinematic fits consistent with the topology of the decay to reconstruct intermediate states and finally $B$ candidates. \par We reconstruct charged-pion candidates using loose requirements on impact parameters (displacement from the nominal interaction space-point, $|dr|<\SI{0.5}{cm}$ radial and $|dz|<\SI{3.0}{cm}$ longitudinal with respect to the beams) to reduce beam-background-induced tracks, which do not originate from the interaction region.
Charged-pion candidates are also restricted to the polar-angle acceptance of the drift chamber ($\SI{17}{\degree}<\theta<\SI{150}{\degree}$).  We reconstruct neutral-pion candidates by combining pairs of photons with energies greater than about $20$\,MeV restricted in diphoton mass to within approximately three times the resolution from the known $\pi^0$ mass~\cite{Zyla:2020zbs} and excluding extreme helicity-angle values  to suppress combinatorial background from collinear low-momentum photons. The mass of the $\pi^0$ candidates is constrained to its known value~\cite{Zyla:2020zbs} in subsequent kinematic fits.
The resulting $\pi^\pm$ and $\pi^0$ candidates are combined to form $\Prhoplus$ and $\Prhozero$ candidates by requiring $0.52 < m(\pi^+\pi^{0,-}) < 1.06$\,GeV/$c^2$. We apply a selection on the cosine of the helicity angle of the $\Prhoplus$ candidates, $\cos\theta_{\rho^+}<0.75$, to suppress candidates formed with low-momentum neutral pions, which usually originate from background.  Pairs of $\Prhoplus$ and $\Prhozero$~candidates are then combined to form $B$-meson candidates. The resulting $B$ candidates are combined through kinematic fits of the entire decay chain. In addition, we reconstruct the vertex of the accompanying tag-side $B$ mesons using  all tracks not associated with the signal $B$ candidate and identify the flavor~\cite{Abudinen:2018}, which is used as  input to the continuum-background discriminator. The reconstruction of the control channels is conceptually similar and includes additional charmed-meson invariant mass restrictions when appropriate.
\par
Simulation is used to identify and suppress contamination from peaking backgrounds, that is, background from $B$-decays clustering in the signal region $M_{\rm bc} > 5.27$\,GeV/$c^2$ and \mbox{$-0.15 < \Delta E < 0.15$ GeV}.  
Peaking backgrounds from decays with intermediate charmed\break 
resonances are suppressed by requiring 
$\vert m(\Pgppm\Pgpmp) - m_{\PDzero}\vert > \SI{18}{MeV}/c^{2}$ and\break 
$\vert m(\Pgppm\Pgpmp\Pgpz) - m_{\PDzero}\vert > \SI{32}{MeV}/c^{2}$, where $m_{\PDzero}$ is the known \PDzero mass~\cite{Zyla:2020zbs}, and\break 
$\vert m(\Pgppm\Pgpmp) - \SI{1.78}{GeV}/c^2\vert > \SI{18}{MeV}/c^{2}$ and 
$\vert m(\Pgppm\Pgpmp\Pgpz) - \SI{1.78}{GeV}/c^2\vert > \SI{32}{MeV}/c^{2}$, where 1.78 GeV/$c^2$ is the peak position of the charm signal when a kaon is misidentified as a pion and ranges correspond to approximately $\pm 1.5$ standard deviations of the relevant mass distribution.


\subsection{Continuum suppression}
The main challenge in observing significant $\PBplus\to \Prhoplus\Prhozero$ signal is the large contamination from continuum background. We use a binary boosted decision-tree classifier ($C'_{
\rm FBDT}$) that combines non-linearly 39 variables known to provide discrimination between $B$-meson signals and continuum and to be loosely correlated, or uncorrelated,  with $\Delta E$ and $M_{\rm bc}$. These variables include quantities associated with event topology (global and signal-only angular configurations), flavor-tagger information, vertex information and uncertainty information, and kinematic-fit quality information. We train the classifier to identify statistically significant signal and background features using simulated samples. 

\begin{figure}[h!]
 \centering
    \centering
    \subfigure{\includegraphics[width=0.49\textwidth]{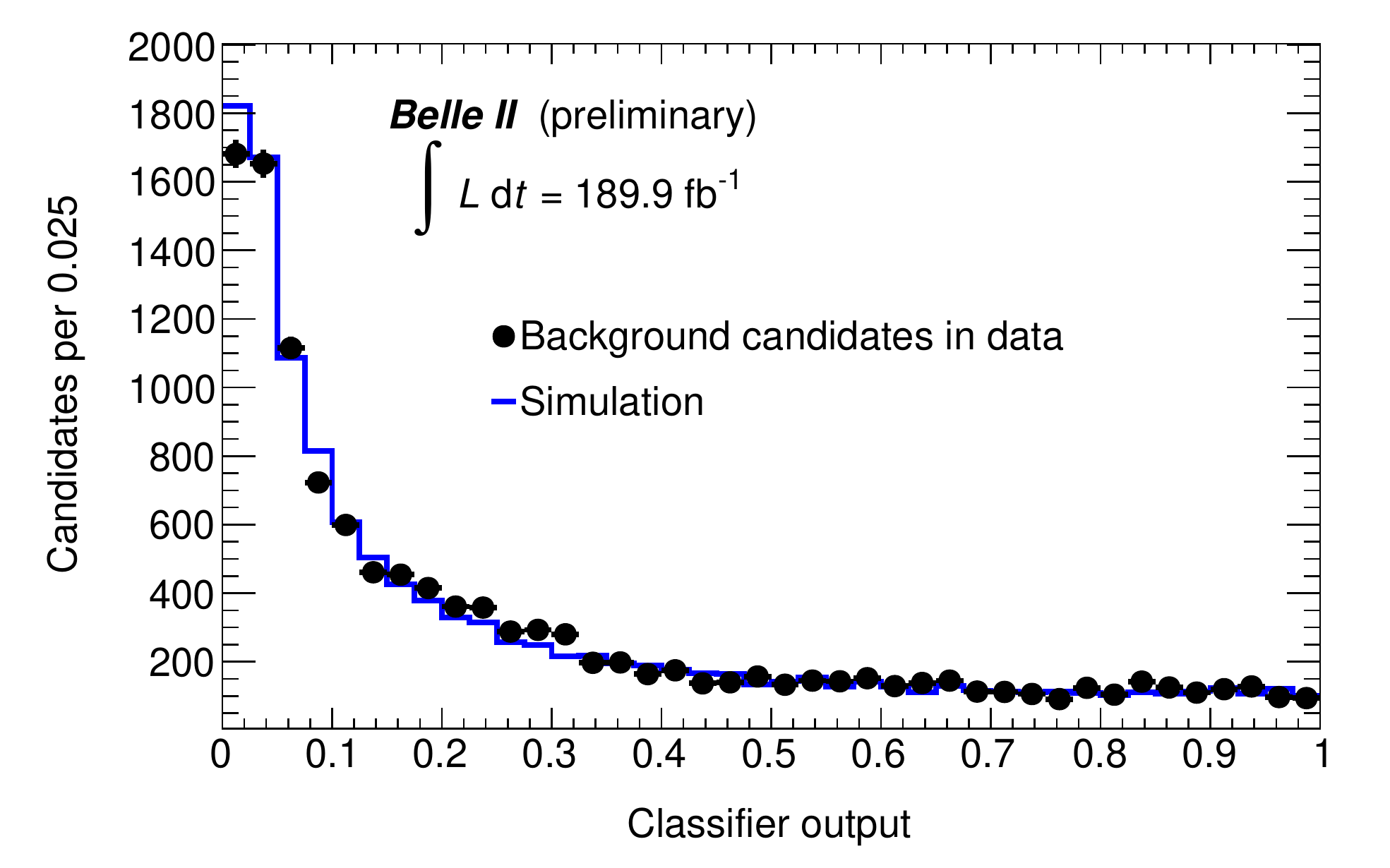}} \hfill
    \subfigure{\includegraphics[width=0.49\textwidth]{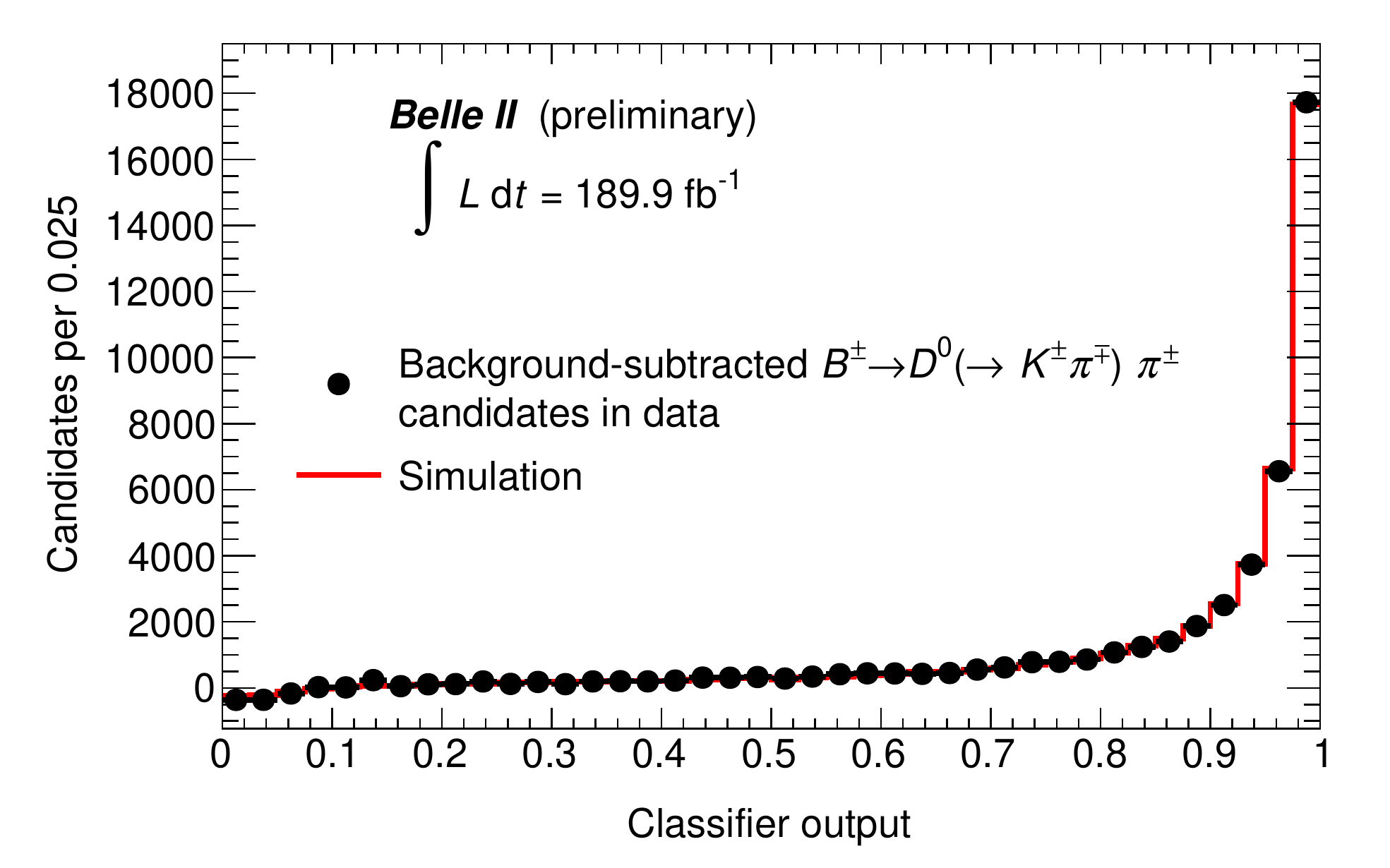}} \\
 \caption{Data-simulation comparison of the output of the boosted decision-tree classifier for (left)~$M_{\rm bc}$~sideband and (right)~$M_{\rm bc}$-sideband-subtracted $\PBplus\to\APD^{0}(\to \PKp\Pgpm)\,\Pgpp$~candidates in the signal region.}
 \label{fig:outputData_Kpi}
\end{figure}

We validate the input and output distributions of the classifier by comparing  data with simulation using control samples.
Figure~\ref{fig:outputData_Kpi} shows the distribution of the output for  \mbox{$\PBplus\to\APD^{0}(\to \PKp\Pgpm)\,\Pgpp$}~candidates. No inconsistency is observed. 


\subsection{Optimization of the signal selection}
\label{sec:yields}
We optimize the selection using simulated and control-sample data. We vary simultaneously the selection criteria on continuum-suppression output and PID information to maximize $S/\sqrt{S+B}$, where $S$ and $B$ are expected signal and background yields, respectively, as estimated in the signal region ($M_{\rm bc} > 5.27$\,GeV/$c^2$ and \mbox{$-0.15 < \Delta E < 0.15$ GeV}) of simulated samples. The optimal selection criterion on the continuum-suppression output removes approximately 98\% of continuum and retains approximately 41\% of signal.
We impose a PID selection for charged pions which  has 96\% efficiency  and 7\% kaon-to-pion misidentification rate. In addition, the $\pi^0$ selection is optimized independently by using control \mbox{$B^+ \to \overline{D}^0(\to K^+\pi^-\pi^0)\pi^+$} decays.

\section{Determination of signal yields}
\label{sec:yields}
More than one candidate per event often populates the selected sample, with an average multiplicity of 1.3. We choose a single candidate per event by selecting the $\pi^0$ candidate with lowest $\chi^2$ value of its mass-constrained diphoton fit, and then by selecting the $B$ candidate with lowest vertex-fit $\chi^2$. The efficiency of these requirements is 92\% (95\%) for the longitudinal (transverse) component.\par Signal yields are determined with likelihood fits of the unbinned distributions of $\Delta E$, $C^\prime_{\rm FBDT}$, $\rho$ candidate dipion masses, and cosines of the helicity angles of the $\rho$ candidates, for candidates in the signal region $M_{\rm bc} > 5.27$\,GeV/$c^2$, $-0.15 < \Delta E < 0.15$ GeV, and $0.52 < m(\pi^+\pi^{-,0}) < 1.06$\,GeV/$c^2$.
Dependencies between the fit observables are modeled by using multidimensional probability density functions (PDFs) of up to three dimensions for each component. 
Sample components used in the fit are transversely and longitudinally polarized signal, $B \to f_0(980)X$ decays, $B \to a_1\pi$ decays, other $B\overline{B}$~background ({\it i.e.}, background from $B$ decays not peaking in $-0.15 < \Delta E < 0.15$ GeV), and continuum, whose yields are determined by the fit. Self cross-feed (i.e., incorrectly reconstructed candidates in signal events) and background from nonresonant $\PB\to\Prho\pi\pi$ decays, are determined by the fit within Gaussian constraints based on expectations from previous measurements~\cite{Zhang:2003up,BaBar:2009rmk} or phenomenological considerations~\cite{Calder_n_2007,PhysRevD.76.079903,PhysRevD.76.114020,Bauer}. Nonresonant $B^+ \to \pi^+\pi^0\pi^+\pi^-$ contributions are assumed to be negligible.\par 
Fit models are physics-based when appropriate (e.g., dipion masses) or empiric otherwise, generally determined by parametrizing distributions from simulation using analytical functions or histogram templates. In particular, multidimensional histogram templates are always used to model dependencies among the fit observables.
We use a sum of three Gaussian functions for all the  $C^\prime_{\rm FBDT}$ shapes and for the peaking background $\Delta E$ shapes. We use a quadratic polynomial for the continuum $\Delta E$ shape and 5th-degree polynomials for the  $m(\pi\pi)$ $\PB\overline{\PB}$~background shapes. All other shapes are modeled with histogram templates from simulation. Corrections to the simulation-based models are taken from fits to the control channel and applied to analytical models and templates.  To account for modest differences between data and simulation, we include a global additive shift of the peak positions and a multiplicative width scale-factor for the $\Delta E$ and $m(\pi^+\pi^0)$ distributions, determined as indicated by likelihood-ratio tests on control channels in data. The properties of the estimators are studied in ensembles of simplified and realistic simulated experiments and found to be satisfactory.\par
\begin{figure}[htb]
 \centering
    \centering
    \subfigure{\includegraphics[width=0.425\textwidth]{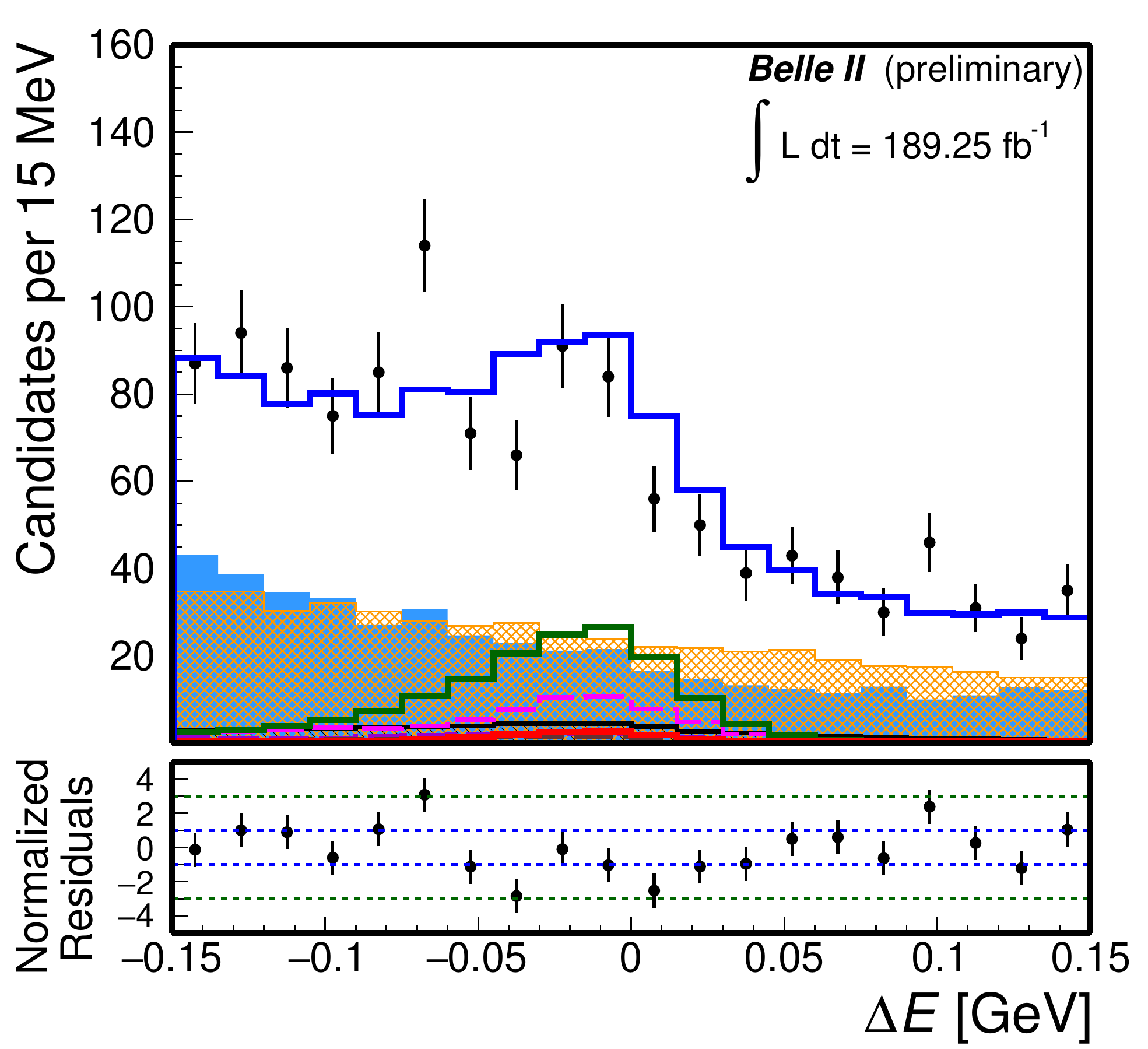}}
    \subfigure{\includegraphics[width=0.425\textwidth]{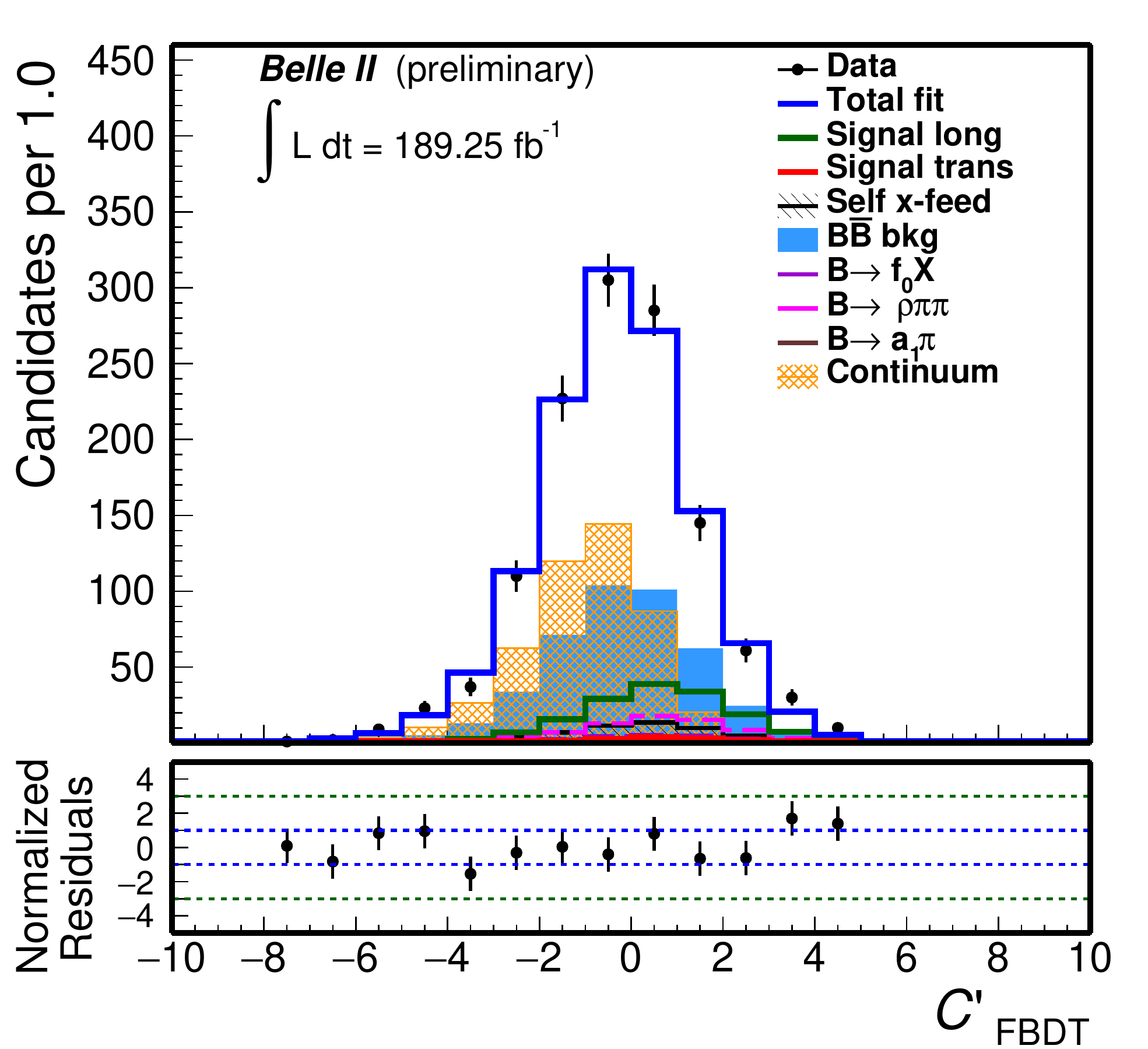}}
    \subfigure{\includegraphics[width=0.425\textwidth]{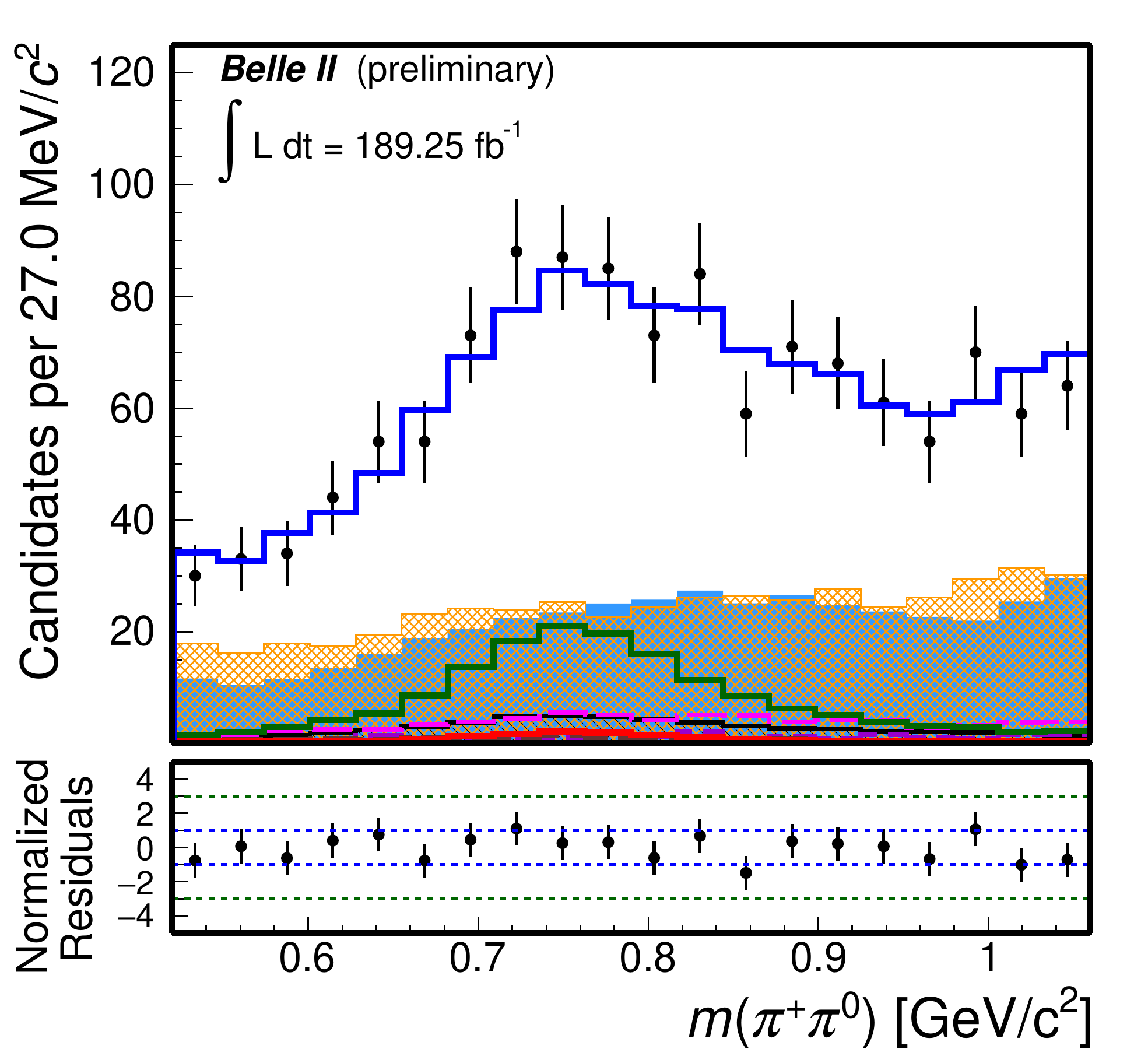}}
    \subfigure{\includegraphics[width=0.425\textwidth]{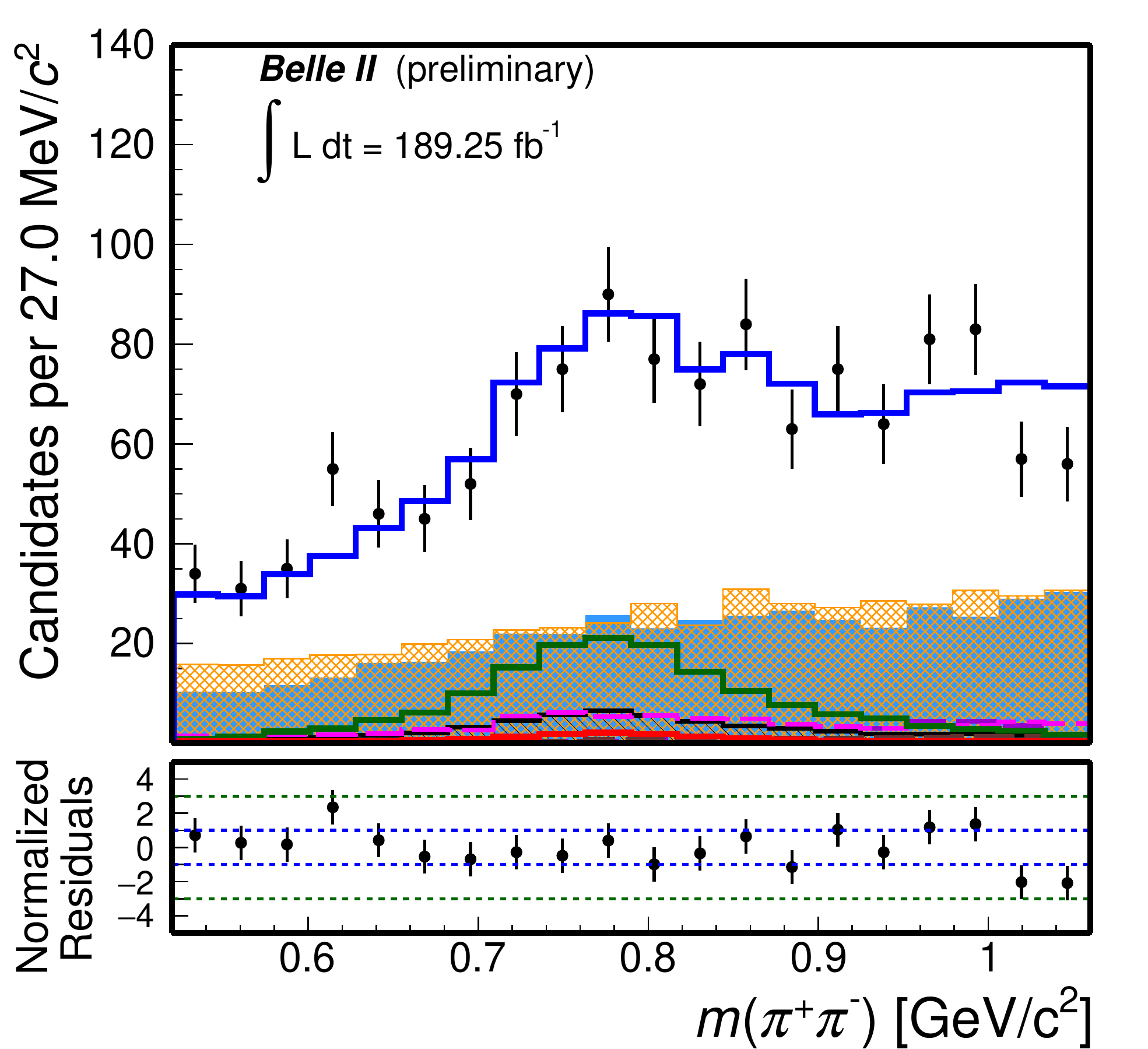}}
    \subfigure{\includegraphics[width=0.425\textwidth]{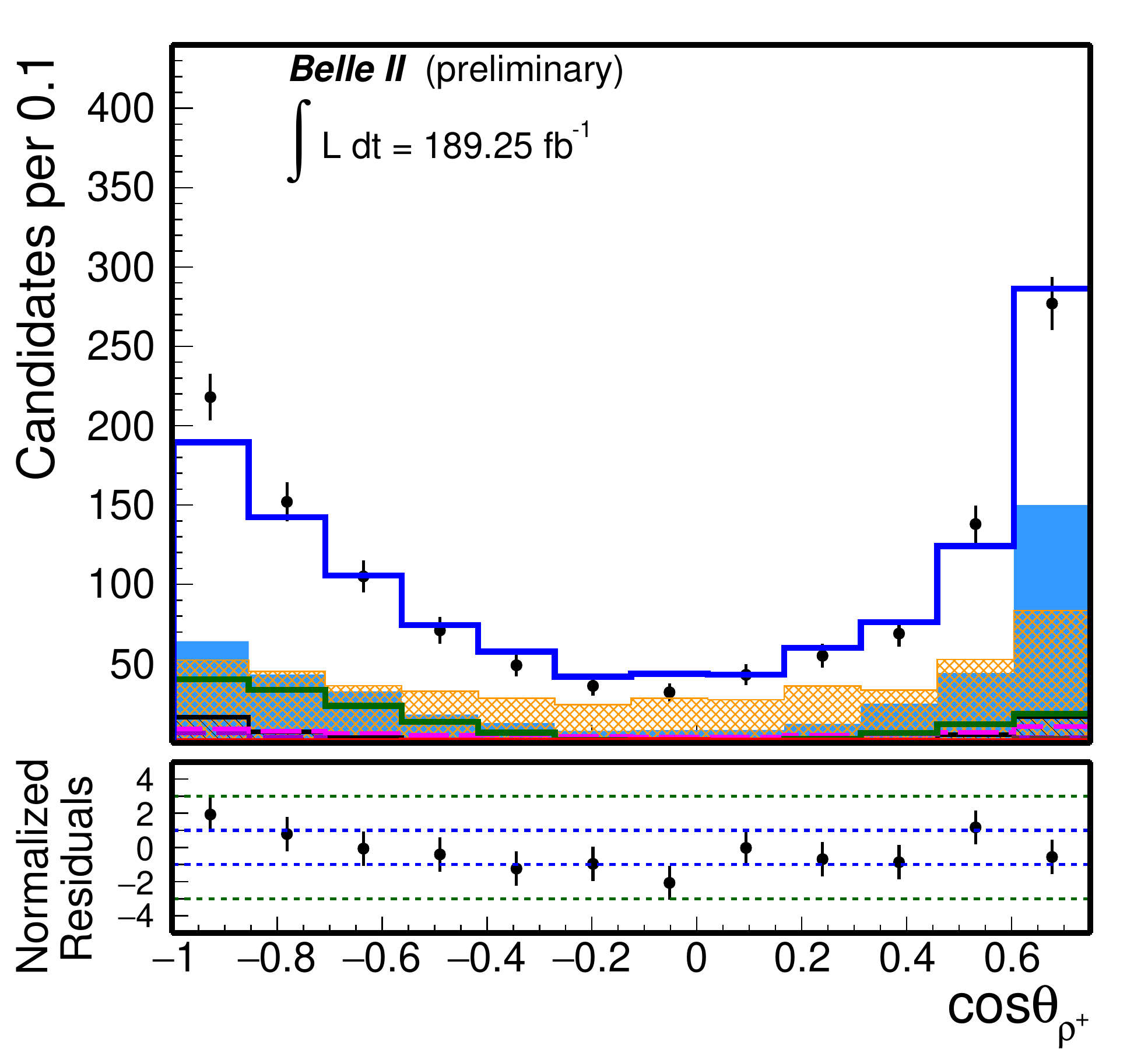}}
    \subfigure{\includegraphics[width=0.425\textwidth]{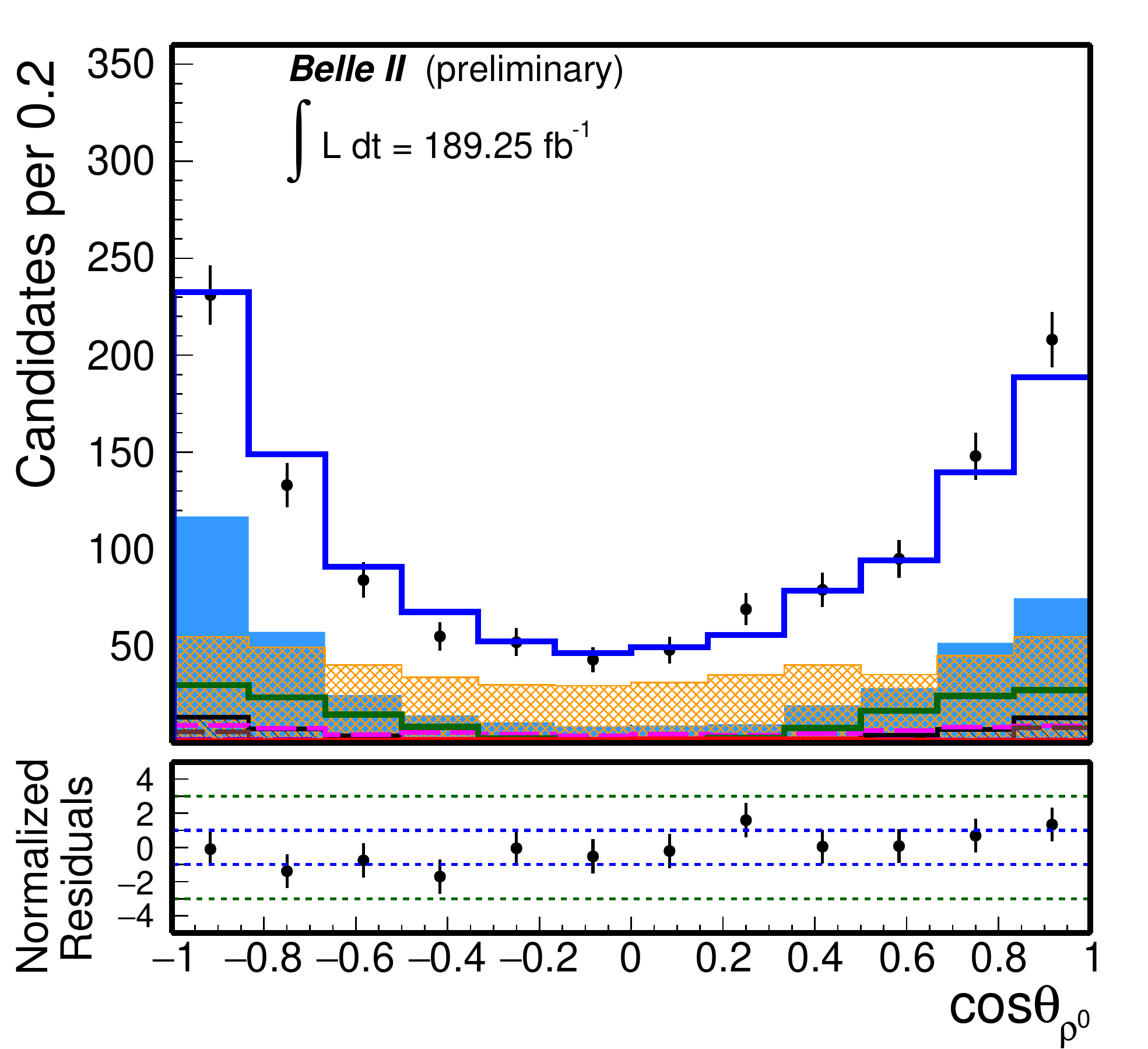}}
 \caption{Distributions of (top left) $\Delta E$, (top right) $C^\prime_{\rm FBDT}$, (middle left) $m(\pi^+\pi^0$), (middle right) $m(\pi^+\pi^-$), cosine of the helicity angle of (bottom left) $\rho^+$ and (bottom right) $\rho^0$ for $B^+ \to \rho^+\rho^0$  candidates (charge-specific) reconstructed in 2019--2021 Belle~II data selected through the baseline criteria with an optimized continuum-suppression and pion-enriching selection, and further restricted to $M_{\rm bc} > 5.27$\,GeV/$c^2$. 
 Fit projections are overlaid.}
 \label{fig:fits_pos}
\end{figure}

\begin{figure}[htb]
 \centering
    \centering
    \subfigure{\includegraphics[width=0.425\textwidth]{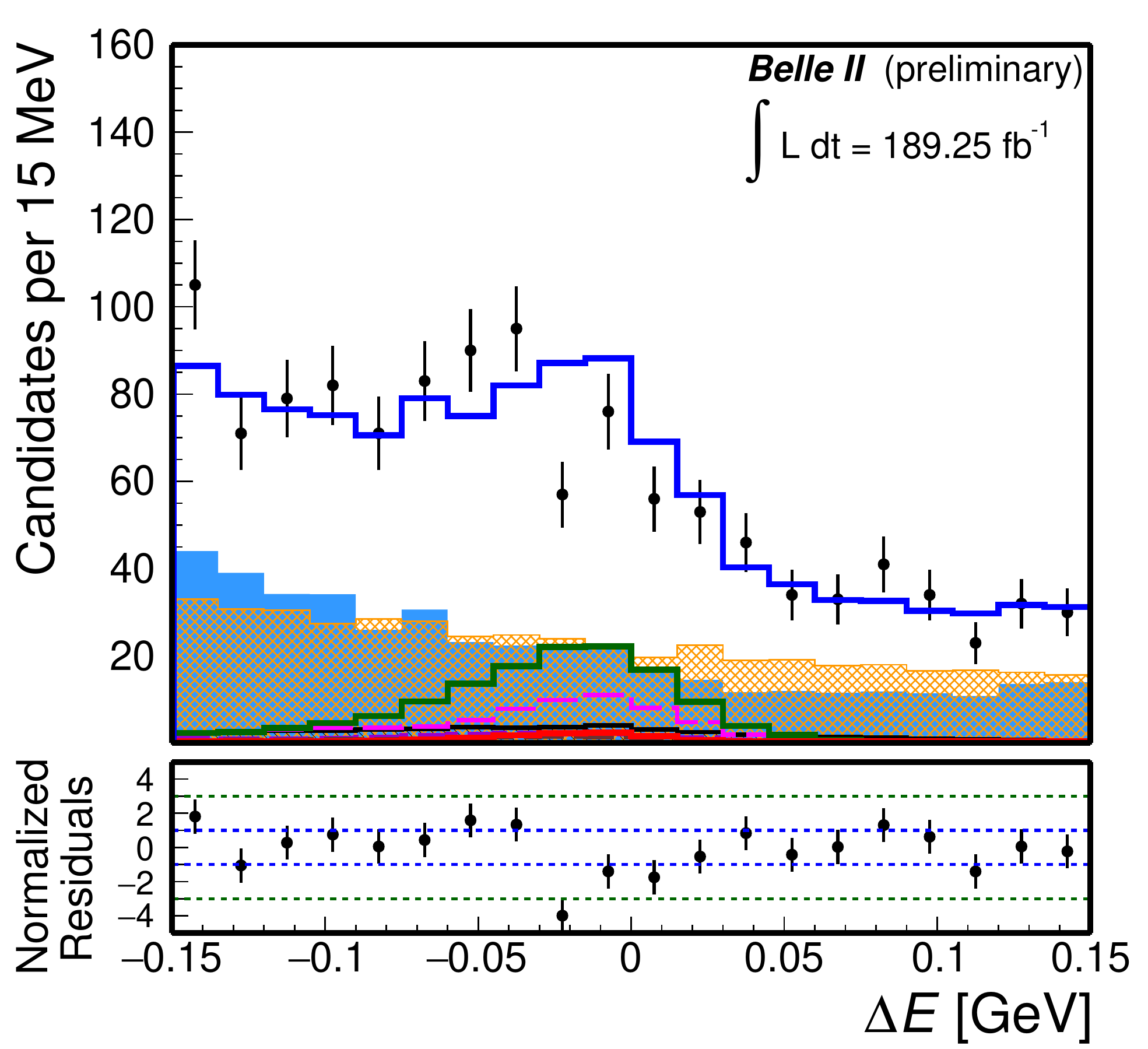}}
    \subfigure{\includegraphics[width=0.425\textwidth]{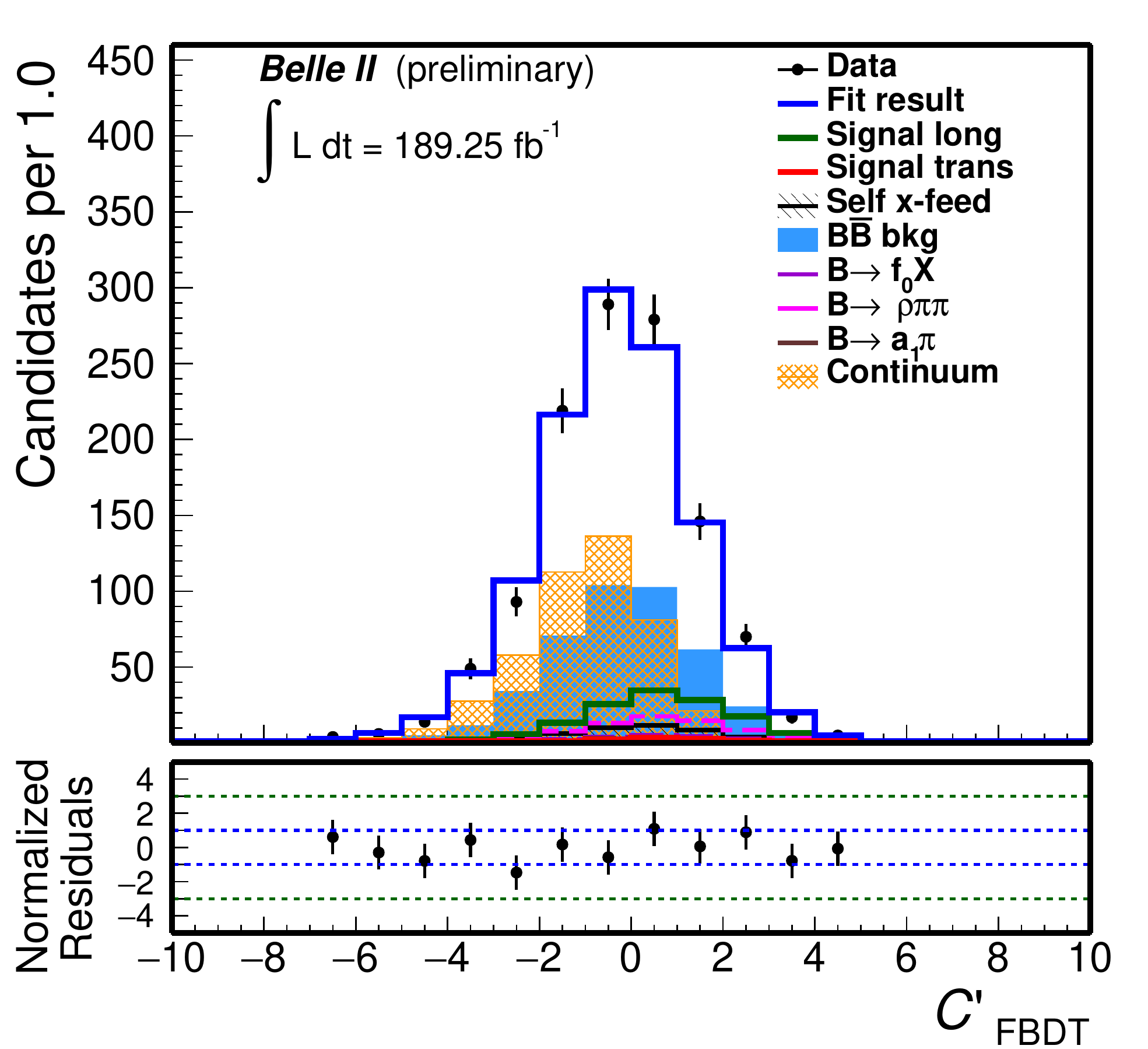}}
    \subfigure{\includegraphics[width=0.425\textwidth]{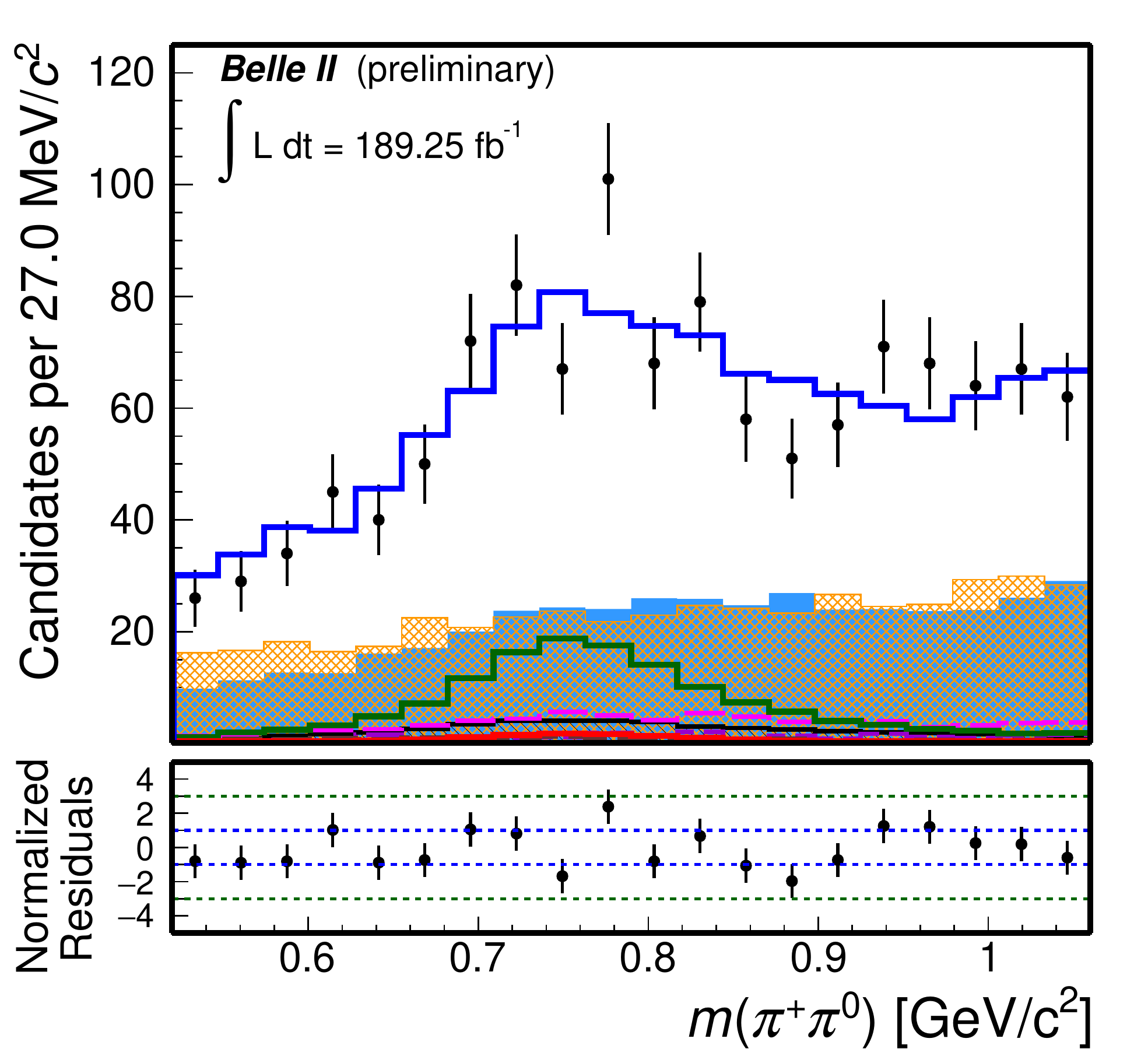}}
    \subfigure{\includegraphics[width=0.425\textwidth]{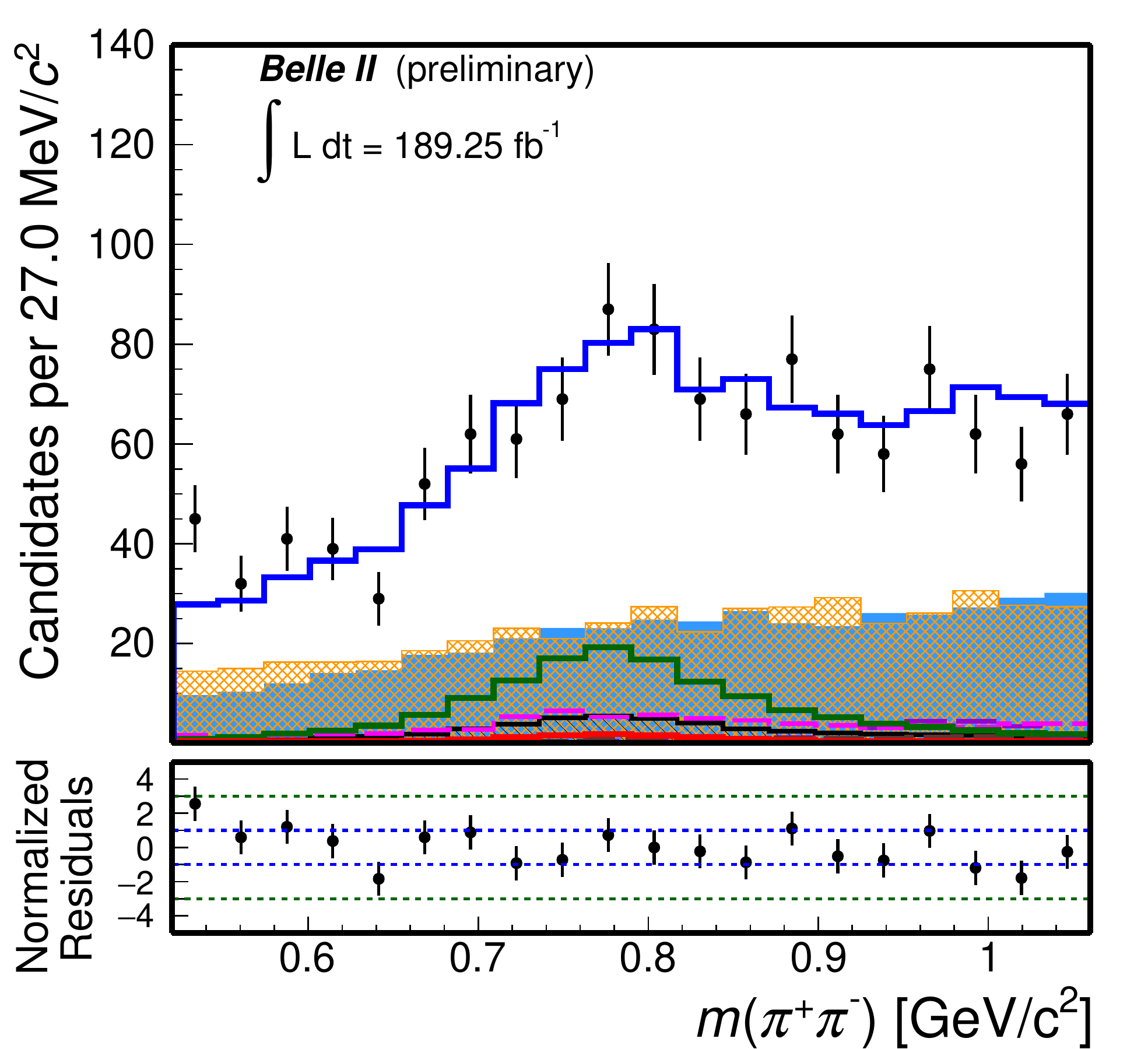}}
    \subfigure{\includegraphics[width=0.425\textwidth]{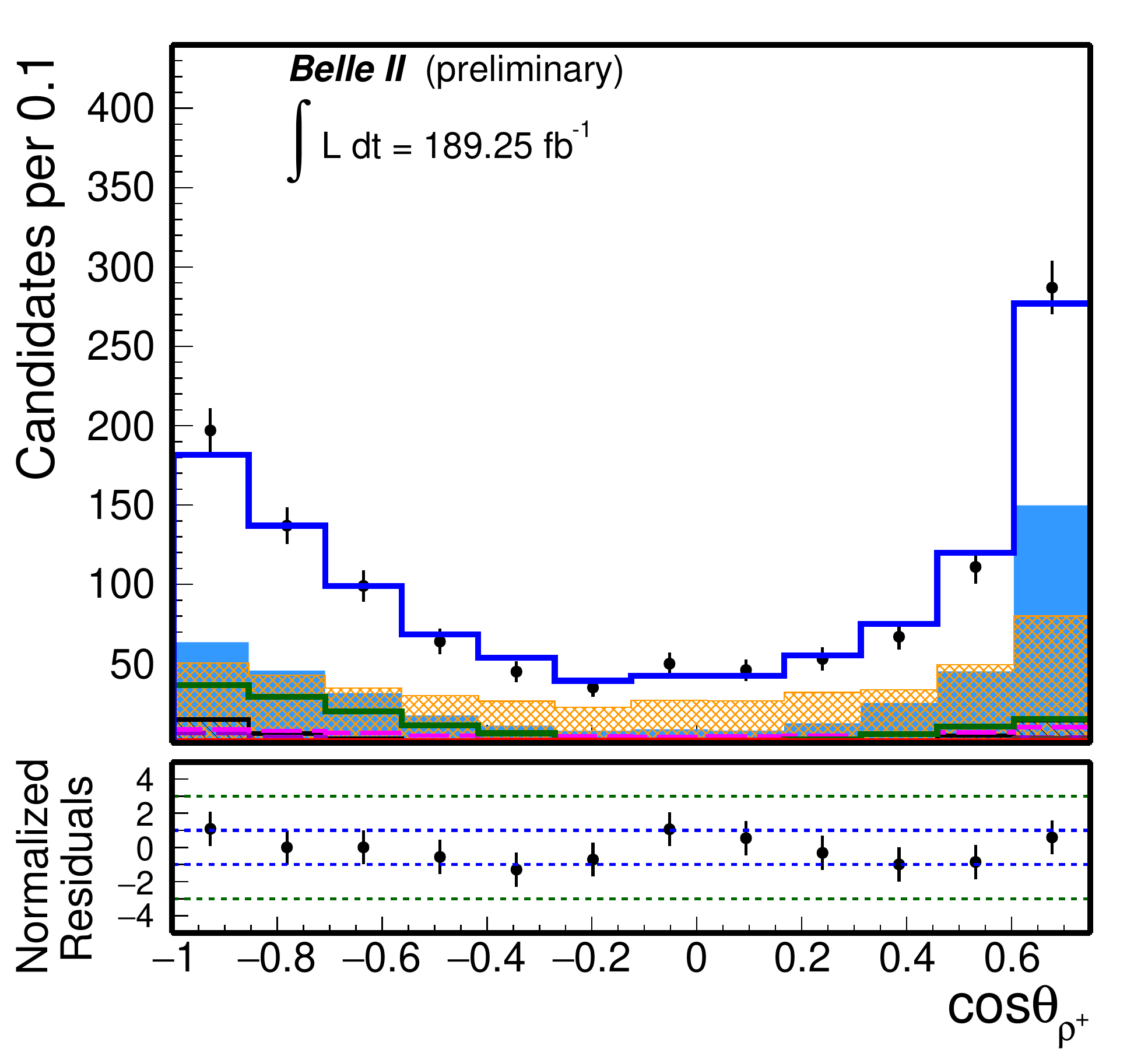}}
    \subfigure{\includegraphics[width=0.425\textwidth]{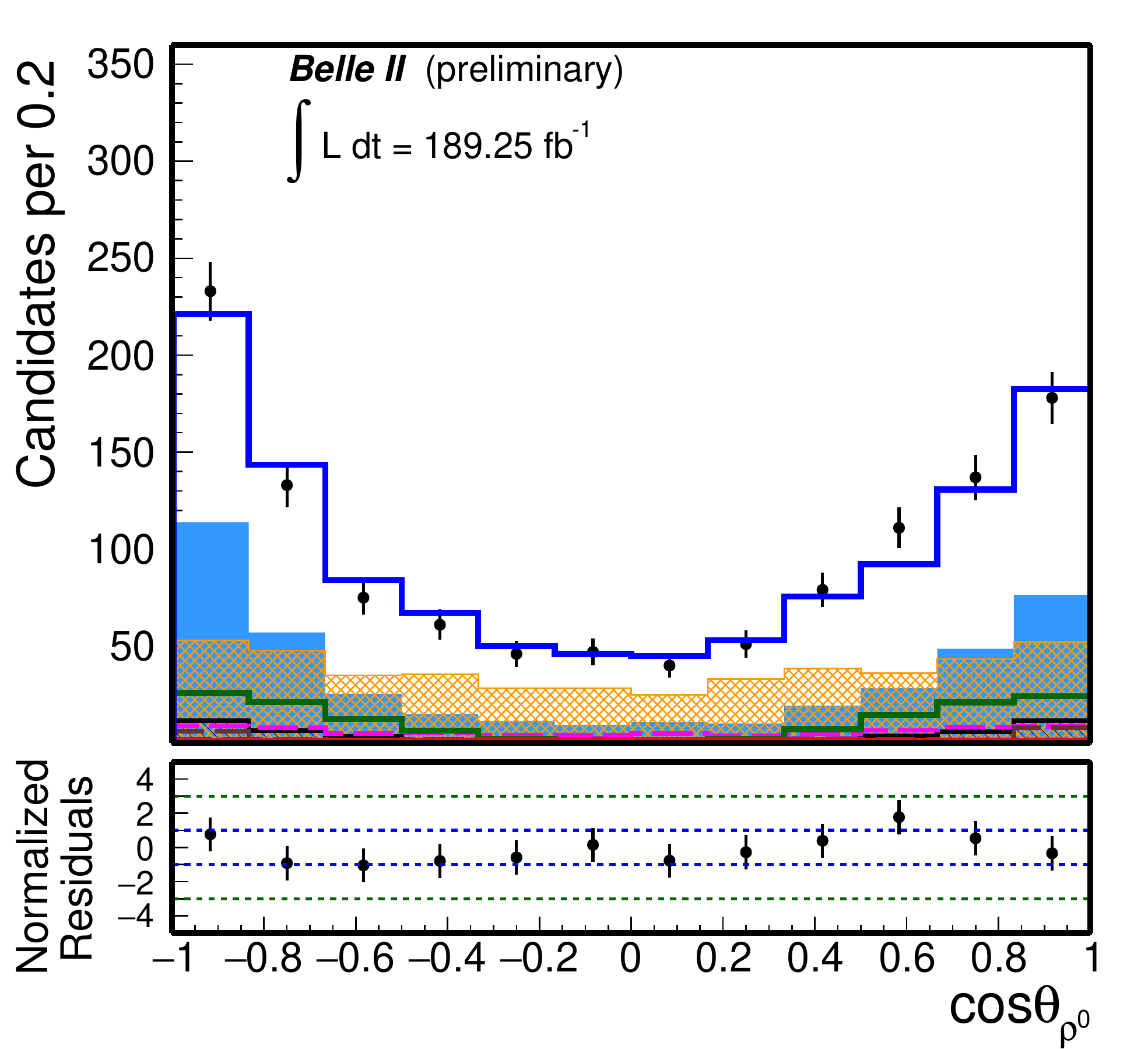}}
 \caption{Distributions of (top left) $\Delta E$, (top right) $C^\prime_{\rm FBDT}$, (middle left) $m(\pi^+\pi^0$), (middle right) $m(\pi^+\pi^-$), cosine of the helicity angle of (bottom left) $\rho^+$ and (bottom right) $\rho^0$ for $B^- \to \rho^-\rho^0$  candidates (charge-specific) reconstructed in 2019--2021 Belle~II data selected through the baseline criteria with an optimized continuum-suppression and pion-enriching selection, and further restricted to $M_{\rm bc} > 5.27$\,GeV/$c^2$. 
 Fit projections are overlaid.}
 \label{fig:fits_neg}
\end{figure}

 
Distributions of fit observables in Belle II data are shown in Figures~\ref{fig:fits_pos} and ~\ref{fig:fits_neg} for positively- and negatively-charged $B$ candidates, respectively, with fit projections overlaid. 
Broadly-peaking signal structures are visible in the energy-difference and dipion mass distributions, otherwise dominated by the smooth continuum and $B\overline{B}$ background components.
The results of the fit are summarized in Table~\ref{tab:FitSummary}.

\begin{table}[!ht]
    \centering
\caption{Fit results in data. Parameters definitions are given in Sects.~\ref{sec:efficiencies} and~\ref{sec:acp}.} 
\begin{tabular}{l  c}
\hline\hline
Parameter & \multicolumn{1}{c}{Fit result} \\\hline
 $f_+$     &  $0.533\pm 0.034$ \\ 
 $\mathcal{B}\ [10^{-6}]$     &  $23.2^{+\ 2.2}_{-\ 2.1}$ \\ 
 $f_{L}$	   &   $0.943 ^{+\ 0.035}_{-\ 0.033}$\\
 Self cross-feed fraction	   &  $0.316\pm 0.028$\\
 $B\overline{B}$ background	   &  $847\pm 51$\\
 $B \to f_0X$ decays &   $51^{+\ 23}_{-\ 22}$	\\
 $B \to \rho\pi\pi$ decays &   $91^{+\ 51}_{-\ 49}$\\
 $B \to a_1\pi$ decays &   $55^{+\ 23}_{-\ 25}$\\
 Continuum background	   &  $939\pm 71$ \\
\hline
\end{tabular}
    \label{tab:FitSummary}
\end{table}{}

\clearpage

%
%

\section{Determination of branching fraction and fraction of longitudinally polarized decays}
\label{sec:efficiencies}

We determine the branching fraction and the fraction of longitudinally polarized decays from the fit, using the relations

\begin{equation*}
    \mathcal{B} = \frac{\frac{N_{L}}{\varepsilon_{L}} + \frac{N_{T}}{\varepsilon_{T}}}{2 N_{B^+B^-} \mathcal{B}_{\pi^0}},\ \ f_L = \frac{\frac{N_{L}}{\varepsilon_{L}}}{\frac{N_{L}}{\varepsilon_{L}} + \frac{N_{T}}{\varepsilon_{T}}}
\end{equation*}
where $N_L$ and $N_T$ are the longitudinally-polarized and transversely-polarized signal yields, respectively;  $\varepsilon_L$ and $\varepsilon_T$ are the corresponding selection and reconstruction efficiencies;  $\mathcal{B}_{\pi^0}$ is  $\mathcal{B}(\pi^0\to\gamma\gamma) = (98.823 \pm 0.034)\%$~\cite{Zyla:2020zbs}; and $N_{B^+B^-} = 101.4\times 10^6$ is the number of produced $\PBplus\PBminus$ pairs. The efficiencies, $\varepsilon_{L}=6.6\%$ and $\varepsilon_{T} = 11.6\%$, include detector acceptance and are determined from simulation with 0.1\% statistical uncertainties. We perform consistency checks on control samples of data and assess systematic uncertainties in the above efficiencies where simulation may not accurately describe data.
We obtain $N_{B^+B^-}$ by multiplying the number of $B$-meson pairs counted in the current data set with $R^{+0}/(R^{+0} + 1)$, where $R^{+0} = 1.058 \pm 0.024$ is the ratio between the branching fractions of the decays of the $\Upsilon(4{\rm S})$ meson to pairs of charged and neutral $B$-meson~\cite{Zyla:2020zbs}. We assume that the $\Upsilon(4{\rm S})$ decays exclusively to ${\PB\APB}$~pairs. We assume 100\% for the  \mbox{$\Prhoplus \to \pi^+\pi^0$} and $\Prhozero \to \pi^+\pi^-$ branching fractions.

\section{Determination of the charge asymmetry}
\label{sec:acp}
To measure the  charge asymmetry $\mathcal{A}$, we apply the same model described above to a simultaneous fit of the samples of charge-specific candidates after fixing the total number of events in the two samples, and using the fraction of positively-charged signal candidates as free parameter in the fit,

\begin{equation*}
    f_+ = \frac{{N}_+}{{N}_+ + {N}_-} = \frac{1-\mathcal{A}}{2},
\end{equation*}
where $N_+$ and  $N_-$ are the charge-specific signal yields. The same parameter is also used for the self cross-feed component, while all the other fit components have their positively-charged fraction fixed to $ f_+=1/2$.

The observed charge-specific raw event-yield asymmetry $\mathcal{A}$ arises in general from the combination of genuine \CP-violating effects in the decay dynamics and instrumental asymmetries due to differences in interaction or reconstruction probabilities between particles and antiparticles. Such a combination is additive for small asymmetries,  $\mathcal{A}=\mathcal{A}_{\CP}+\mathcal{A}_{\rm det}$.
Hence, observed raw charge-specific decay yields need be corrected for instrumental effects to determine the genuine \CP-violating asymmetries.

We measure the instrumental asymmetry due the reconstruction of a charged pion using the \mbox{$D^+ \to K^0_S\pi^+$} channel. We subtract the $K^0_S$ contribution from the $K^0_S\pi$ asymmetry through the relationship \mbox{$\mathcal{A}(\pi^+)=\mathcal{A}(K^0_S\pi^+)-\mathcal{A}(K^0_S)$}, where the direct \CP asymmetry in \mbox{$D^+ \to K^0_S \pi^+$} decays is subtracted and the contribution deriving from the kaon {\CP} asymmetry $\mathcal{A}(K^0_S)$ is estimated by using the results obtained by the LHCb experiment~\cite{LHCbInstr:2018}; we assume $\mathcal{A}(K^0_S)$ to be zero with an associated uncertainty of $0.2\%$. This approach is consistent with the one at Belle, where $\mathcal{A}(K^0_S)$ was assumed to be negligible~\cite{Ko:2010mk}.
We find a value of
\begin{equation}
\begin{aligned}
\mathcal{A}_{\rm det}(\pi^+) = +0.0040 \pm 0.0048.
\end{aligned}
\label{eq:InstrAsym}
\end{equation}
We detail the determination of the instrumental asymmetries in Ref.~\cite{CharmlessICHEP:2020}.

\section{Systematic uncertainties}

We consider several systematic effects, assumed to be independent, and add in quadrature the corresponding uncertainties. An overview of the effects considered follows. A summary of the systematic uncertainties is listed in Table~\ref{tab:Systematics_overview}.

\subsection{Tracking efficiency}
We assess a systematic uncertainty associated with possible data-simulation discrepancies in the reconstruction of charged particles~\cite{Bertacchi:2020eez}.
The tracking efficiency in data agrees with the value observed in simulation within a $0.30\%$ uncertainty, which we (linearly) add as a systematic uncertainty for each final-state charged particle.

\subsection{\Pgpz~reconstruction efficiency} 
We assess a systematic uncertainty associated with possible data-simulation discrepancies in the $\Pgpz$ reconstruction and selection using the decays \mbox{$B^0 \to D^{*-}(\to \overline{D}^0 (\to K^+ \pi^- \pi^0)\, \pi^-)\, \pi^+$} and \mbox{$B^0 \to D^{*-}(\to \overline{D}^0 (\to K^+ \pi^-)\, \pi^-)\, \pi^+$} where the selection of charged particles is identical and the $\pi^0$ momentum is similar to the momentum in $B^+ \to \rho^+\rho^0$. We compare the yields obtained from fits to the $\Delta E$ distribution of reconstructed $\PB$~candidates and obtain an efficiency in data that agrees with that observed in simulation. The ratio is compatible with unity within an approximate 6.5\% uncertainty, which is used as systematic uncertainty.

\subsection{Particle-identification and continuum-suppression efficiencies} 
We identify possible data-simulation discrepancies in PID and continuum-suppression distributions using the control channel  \mbox{$\PBplus\to\APDzero(\to\PKp\Pgpm)\,\Pgpp$}. The ratio between the efficiencies in data and simulation,  $R = 0.960 \pm 0.012$, is incompatible with unity. We scale the $\mathcal{B}$ result by this ratio and take the uncertainty as a systematic uncertainty associated with the PID and continuum-suppression selection efficiencies.

\subsection{Number of $B^+B^-$~pairs} We  assign  a  systematic  uncertainty associated with the uncertainty on the number of $B^+B^-$~pairs,  which  includes  the uncertainty  on  cross-sections,  integrated  luminosity, and  potential  shifts  from  the peak center-of-mass energy during the run periods.

\subsection{Signal and background modeling} We quote, as the systematic uncertainty for possible signal or background mismodelings, the difference between the average of results of the sample-composition fit performed on ensembles of  simplified simulated experiments generated with the baseline model and various alternate models. Uncertainties on branching fraction, longitudinal polarization fraction, and charge-parity asymmetry range from 0.02\% to 1.20\%.

\subsection{Single candidate selection}

We assess a systematic uncertainty on the single-candidate selection by re-doing the analysis selecting randomly one $\PBplus$ candidate per event. The difference between results is taken as a systematic uncertainty.

\subsection{Fit biases}

We include in the systematic uncertainty statistical biases observed in fits of ensembles of simplified simulated experiments. 

\subsection{Simulation}

The largest systematic uncertainty is due to significant data-simulation discrepancies observed in distributions of fit observables for candidates populating signal sidebands, which could not be conclusively attributed to shape or acceptance mismodelings or poorly simulated sample composition. A systematic uncertainty based on the deviation of results in fits to simulated ensembles that mirror the observed discrepancies generously covers any possible effect. 

\subsection{Instrumental asymmetries}
We consider the uncertainties on the values of  $\mathcal{A}_{\rm det}$~(Eq.~\ref{eq:InstrAsym}) as systematic uncertainties due to instrumental asymmetry corrections in measurements of \CP~asymmetries.

\subsection{Peaking background \CP asymmetries}
We assume continuum and $B\bar{B}$ background components to be charge-symmetric, as supported by fits in sidebands. To account for possible \CP violation in the peaking background, we generate simplified simulated experiments with nonzero \CP asymmetry for the $\rho\pi\pi$ component and fit them with the baseline model. The average values of the residuals of these fits are taken as systematic uncertainties, which are 0.046 for $\mathcal{A}_{\CP}$ and sub-percent for $\mathcal{B}$ and $f_L$.

\begin{table}[h]
\centering
\footnotesize
\caption{Summary of the (fractional) systematic uncertainties of the branching-fraction and longitudinal-polarization-fraction measurements.}
\begin{tabular}{l c c c}
\hline\hline
 Source & $\mathcal{B}$ & $f_L$ & $\mathcal{A}_{\CP}$\\
\hline
Tracking             & 0.9\% & n/a & n/a\\
$\Pgpz$ efficiency      & 5.7\%  & n/a & n/a\\
PID and continuum-supp. eff. & 1.2\% & n/a & n/a\\
$N_{B^+B^-}$      &  3.1\%  &  n/a & n/a\\
Instrumental asymmetry correction & n/a & n/a & 0.005\\
Single candidate selection & 2.2\% & 1.1\% & 0.037\\
Signal model & 0.10\%  & 0.02\% & 0.002\\
Continuum bkg.\ model &  0.04 \%  & 1.2\% & 0.003\\
$\PB\APB$ bkg.\ model &  0.05\%  & 0.08\%  & 0.002\\
Fit biases & 4.4\% & 1.1\%  & 0.010\\
Data-simulation mismodeling & 8.0\% & 2.1\% & 0.002\\
Peaking background \CP asymmetries & 0.3\% & 0.1\% & 0.046\\
\hline
Total  & 11.5\%  & 2.9\%  & 0.060 \\
\hline\hline
\end{tabular} 

\label{tab:Systematics_overview}
\end{table}

\clearpage
\section{Results and summary}
\label{sec:summary}
We report on a measurement of the branching fraction, longitudinal polarization fraction, and \CP~asymmetry of $B^+\to \rho^+\rho^0$ decays. We use a sample of 2019, 2020, and 2021 data collected by the Belle II experiment and corresponding to $190$\,fb$^{-1}$ of integrated luminosity. We use simulation to determine optimized event selections
and fit the $\Delta E$, continuum-background suppression, invariant masses, and angular distributions of the resulting samples to determine a $B^+\to \rho^+\rho^0$ signal yield of $345 \pm 31$
decays. The signal yields are corrected for efficiencies determined from simulation and control data samples to obtain 
\begin{center}
$\mathcal{B}(B^+ \to \rho^+\rho^0) = [23.2^{+\ 2.2}_{-\ 2.1} (\rm stat) \pm   2.7 (\rm syst)]\times 10^{-6}$,
\end{center}
\begin{center}
$f_L = 0.943 ^{+\ 0.035}_{-\ 0.033} (\rm stat)\pm 0.027(\rm syst)$,
\end{center}
\begin{center}
$\mathcal{A}_{\CP}=-0.069 \pm 0.068(\rm stat) \pm 0.060 (\rm syst)$.
\end{center}
This is the first measurement of the \CP~asymmetry of $\PBplus\to\Prhoplus\Prhozero$ decays reported by Belle~II. 
The results on branching fraction and longitudinal polarization fraction presented supersede the previous Belle~II results~\cite{Belle-II:2021rhoprho0}, which were based on a smaller dataset.
Results are compatible with previous determinations~\cite{Zhang:2003up,BaBar:2009rmk} and show performance superior to early Belle results~\cite{Zhang:2003up}.

\clearpage

\section*{Acknowledgments}

We thank the SuperKEKB group for the excellent operation of the
accelerator; the KEK cryogenics group for the efficient
operation of the solenoid; the KEK computer group for
on-site computing support.

\bibliography{belle2}

\providecommand{\href}[2]{#2}\begingroup\raggedright\begin{thebibliography}{10}

\bibitem{Kou:2018nap}
W.~Altmannshofer {\it et al.} {(Belle~II Collaboration)}, { {The Belle II
  Physics Book}\/},  \href{http://dx.doi.org/10.1093/ptep/ptz106,
  10.1093/ptep/ptaa008}{\color{blue}PTEP {\bf 2019} (2019)  123C01}.

\bibitem{Gronau:1990ka}
M.~Gronau and D.~London{{}}, { {Isospin analysis of CP asymmetries in $B$
  decays}\/},
  \href{http://dx.doi.org/10.1103/PhysRevLett.65.3381}{\color{blue}Phys. Rev.
  Lett. {\bf 65} (1990)  3381}.

\bibitem{Abe:2010sj}
T.~Abe {\it et al.} {(Belle II Collaboration)}, { {Belle II Technical Design
  Report, (2010)}\/},   \href{http://arxiv.org/abs/1011.0352}{{\color{blue} \tt
  arXiv:1011.0352}}.

\bibitem{Akai:2018mbz}
K.~Akai {\it et al.} {(SuperKEKB Accelerator Team)}, { {SuperKEKB Collider}\/},
   \href{http://dx.doi.org/10.1016/j.nima.2018.08.017}{\color{blue}Nucl.\
  Instrum.\ Meth.\ A {\bf 907} (2018)  188}.

\bibitem{Lange:2001}
D.~J. Lange{{}}, { {The EvtGen particle decay simulation package}\/},
  \href{http://dx.doi.org/https://doi.org/10.1016/S0168-9002(01)00089-4}{\color{blue}Nucl.\
  Instrum.\ Meth.\ A {\bf 462} (2001)  152}.

\bibitem{Kuhr:2018lps}
T.~Kuhr {\it et al.} { {The Belle II Core Software}\/},
  \href{http://dx.doi.org/10.1007/s41781-018-0017-9}{\color{blue}Comput. Softw.
  Big Sci. {\bf 3} (2019)  1}.

\bibitem{Zyla:2020zbs}
P.~Zyla {\it et al.} , {Particle Data Group}, { {Review of Particle
  Physics}\/},  \href{http://dx.doi.org/10.1093/ptep/ptaa104}{\color{blue}PTEP
  {\bf 2020} (2020) no.~8, 083C01}.

\bibitem{Abudinen:2018}
F.~Abudin{\'e}n{, Ph.D. Thesis}, { {Development of a $\PBzero$~flavor tagger
  and performance study of a novel time-dependent $\CP$ analysis of the decay
  $\PBzero\to\Pgpz\Pgpz$ at Belle~II, Ludwig Maximilian University of Munich
  (2018)}\/},  \href{https://docs.belle2.org/record/1215?ln=en,
  }{\color{blue}BELLE2-PTHESIS-2018-003}.

\bibitem{Zhang:2003up}
J.~Zhang {\it et al.} {(Belle Collaboration)}, { {Observation of $B^+\to \rho^+
  \rho^0$}\/},
  \href{http://dx.doi.org/10.1103/PhysRevLett.91.221801}{\color{blue}Phys. Rev.
  Lett. {\bf 91} (2003)  221801}.

\bibitem{BaBar:2009rmk}
B.~Aubert {\it et al.} , {(BaBar collaboration)}, { {Improved Measurement of
  $B^+ \to \rho^+\rho^0$ and Determination of the Quark-Mixing Phase Angle
  $\alpha$}\/},
  \href{http://dx.doi.org/10.1103/PhysRevLett.102.141802}{\color{blue}Phys.
  Rev. Lett. {\bf 102} (2009)  141802}.

\bibitem{Calder_n_2007}
G.~Calderón, J.~H. Muñoz, and C.~E. Vera, { Nonleptonic two-body B decays
  including axial-vector mesons in the final state\/},
  \href{http://dx.doi.org/10.1103/physrevd.76.094019}{\color{blue}Phys. Rev. D
  {\bf 76} (2007)  }.

\bibitem{PhysRevD.76.079903}
V.~Laporta, G.~Nardulli, and T.~N. Pham, { Nonleptonic $B$ decays to
  axial-vector mesons and factorization [Phys. Rev. D 74, 054035 (2006)]\/},
  \href{http://dx.doi.org/10.1103/PhysRevD.76.079903}{\color{blue}Phys. Rev. D
  {\bf 76} (2007)  079903}.

\bibitem{PhysRevD.76.114020}
H.~Y. Cheng and K.-C. Yang, { Hadronic charmless $B$ decays
  $B\ensuremath{\rightarrow}AP$\/},
  \href{http://dx.doi.org/10.1103/PhysRevD.76.114020}{\color{blue}Phys. Rev. D
  {\bf 76} (2007)  114020}.

\bibitem{Bauer}
M.~Bauer, B.~Stech, and M.~Wirbel, { Exclusive non-leptonic decays of $D^-$,
  $D_s^-$ and $B^-$ mesons\/},
  \href{http://dx.doi.org/10.1007/BF01561122}{\color{blue}Z. Phys. C -
  Particles and Fields {\bf 34} (1987)  103--115}.

\bibitem{LHCbInstr:2018}
A.~Davis {\it et al.} {(LHCb Collaboration)}, { {Measurement of the
  instrumental asymmetry for \mbox{$K^-\pi^+$}-pairs at LHCb in Run~2}\/},
  \href{https://cds.cern.ch/record/2310213/files/LHCb-PUB-2018-004.pdf,
  }{\color{blue}LHCb-PUB-2018-004}.

\bibitem{Ko:2010mk}
B.~R. Ko, E.~Won, B.~Golob, and P.~Pakhlov, { {Effect of nuclear interactions
  of neutral kaons on CP asymmetry measurements}\/},
  \href{http://dx.doi.org/10.1103/PhysRevD.84.111501}{\color{blue}Phys. Rev. D
  {\bf 84} (2011)  111501},
  \href{http://arxiv.org/abs/1006.1938}{{\color{blue} \tt arXiv:1006.1938
  [hep-ex]}}.

\bibitem{CharmlessICHEP:2020}
F.~Abudin{\'e}n {\it et al.} {(Belle II Collaboration)}, { {Measurements of
  branching fractions and CP-violating charge asymmetries in charmless
  $B$-decay reconstruction in 2019--2020 Belle~II data}\/},
  \href{http://arxiv.org/abs/2009.09452}{{\color{blue} \tt arXiv:2009.09452}}.

\bibitem{Bertacchi:2020eez}
V.~Bertacchi {\it et al.} {(Belle II Tracking Group)}, { {Track Finding at
  Belle II}\/},   (2020)  ,
  \href{http://arxiv.org/abs/2003.12466}{{\color{blue} \tt arXiv:2003.12466}}.

\bibitem{Belle-II:2021rhoprho0}
F.~Abudin\'en {\it et al.} , {Belle-II}, { {Angular analysis of $B^+ \to
  \rho^+\rho^0$ decays reconstructed in 2019-2020 Belle II data}\/},
  \href{http://arxiv.org/abs/2109.11456}{{\color{blue} \tt arXiv:2109.11456
  [hep-ex]}}.

\end{thebibliography}\endgroup
\bibliographystyle{belle2-note}

\end{document}